\documentclass{article}
\usepackage{spconf,amsmath,graphicx}
\usepackage{booktabs,arydshln}
\usepackage{siunitx}

\title{Enhance audio generation controllability through representation similarity regularization}
\name{%
\begin{tabular}{@{}c@{}}
Yangyang Shi \qquad 
Gael Le Lan \qquad 
Varun Nagaraja \qquad
Zhaoheng Ni \qquad
Xinhao Mei \\
Ernie Chang \qquad 
Forrest Iandola \qquad 
Yang Liu \qquad
Vikas Chandra 
\end{tabular}}
\address{Meta AI}
\begin{document}
%
\maketitle
\begin{abstract}
This paper presents an innovative approach to enhance control over audio generation by emphasizing the alignment between audio and text representations during model training. In the context of language model-based audio generation, the model leverages input from both textual and audio token representations to predict subsequent audio tokens. However, the current configuration lacks explicit regularization to ensure the alignment between the chosen text representation and the language model's predictions. Our proposal involves the incorporation of audio and text representation regularization, particularly during the classifier-free guidance (CFG) phase, where the text condition is excluded from cross attention during language model training. The aim of this proposed representation regularization is to minimize discrepancies in audio and text similarity compared to other samples within the same training batch. Experimental results on both music and audio generation tasks demonstrate that our proposed methods lead to improvements in objective metrics for both audio and music generation, as well as an enhancement in the human perception for audio generation.

\end{abstract}
\section{Introduction}
\label{sec:intro}
Generating sound effects, music, and speech to meet specific requirements holds immense importance as a pivotal tool in content creation spanning various domains, including augmented, virtual and mixed reality, video game development, and movie production. The advent of recent neural generative models have brought about a transformative shift in the landscape of digital content generation. Drawing inspiration from the remarkable progress in image generation~\cite{Rombach2022-yo,Ramesh2022-sn}, the realm of audio generation has undergone a paradigm shift – transitioning from conventional signal processing approaches to neural generative models ~\cite{Song2020-go,Liu2023-uz,Liu2023-bg,Kreuk2022-pc,Agostinelli2023-rc,Copet2023-sx,Le2023-fx,Lam2023-lu}.

Just as in the case of text-to-image generation models~\cite{Rombach2022-yo,Dhariwal2021-tg}, harnessing the potential of diffusion probability models~\cite{Ho2020-pp,Kingma2021-zs}, the studies~\cite{Le2023-fx,Huang2022-sc,Kim2022-xl,Shen2023-wp,Liu2023-uz,Liu2023-bg,Huang2023-zw,Schneider2023-ju} have showcased impressive capacity in the realms of speech synthesis, sound effects creation, and music generation. Alongside the diffusion-based approach, a parallel avenue has been pursued using transformer-based language models~\cite{Vaswani2017-mw}, which have also exhibited exceptional performance in audio generation tasks~\cite{Borsos2023-xo,Dunbar2021-os,Lakhotia2021-ir,Copet2023-sx,Kreuk2022-pc,Agostinelli2023-rc}. 

In language model driven approach like MusicGen~\cite{Copet2023-sx} and AudioGen~\cite{Kreuk2022-pc}, it first encodes raw audio into discrete tokens via a neural audio compression model (e.g.,~\cite{Zeghidour2022-dr,Defossez2022-ck}). This model is end-to-end trained to compress and reconstruct input audio from discrete tokens with high quality and minimum perceptual loss. The generation model then employs an auto regressive transformer-decoder language model. The language model operates on discrete audio tokens from the first phase and is conditioned on text inputs. Text is processed as text embedding representation using an text encoder pretrained on a large text corpus, such as T5~\cite{Raffel2019-ba}. The text representation is used as cross attentions in the language model training. The language model is trained by cross-entropy loss to minimize the entropy to predict next discrete audio token based on the previous audio tokens and the text representation. However, in the whole training process, there is not any regularization to enforce the next audio token prediction to fully leverage representations from both audio token and conditioning text. As a consequence, the generated audio often isn't fully aligned with the provided text prompt. It is often that the music generated based on the description "\textit{Highly rhythmic orchestral piece illustrating wonder and awe. Features staccato violins, cellos, basses, trombone and grand piano}",  misses one or more instruments from the description. The sound effects generated from the condition "\textit{the sound of a ping pong ball bounce back once from the hard wood floor}" has multiple ping pong ball bouncing sounds.

This paper introduces a method aiming at improving the training of the generation model to effectively capture representations from text conditions. This is achieved by minimizing the similarity between text and audio representations through regularization. Language model training comprises two modes: text-conditioned training and classifier-free guidance (CFG) training~\cite{Ho2022-wm,Kreuk2022-pc}. In CFG, the text condition is omitted during language model training. We enhance the audio and text representation similarity by reducing discrepancies in audio and text similarity compared to other samples within the same training batch. Experimental results in music and sound effects generation demonstrate the effectiveness of the proposed approach, showcasing improvements in Frechet audio distance (FAD) using VGG classifier~\cite{Hershey2017-fk}, kullback–leibler (KL) divergence using PaSST model~\cite{Koutini2021-fx}, text and audio alignment score based on the contrastive language audio pretrained models (CLAP)~\cite{Elizalde2022-ge}, and human subjective evaluation for audio generation.

\vspace{-3mm}
\section{Related Work}
\label{sec:related_work}
This study applies the language model approach presented in works such as~\cite{Borsos2023-xo,Dunbar2021-os,Lakhotia2021-ir,Copet2023-sx,Kreuk2022-pc,Agostinelli2023-rc}, in which the compression model discretizes audio into tokens for training and then decodes these tokens to audio. The language model learns to generate audio tokens. However, our emphasis lies in augmenting the semantic correlation between provided text descriptions and the generated audio. This enhancement is built upon the foundation of the MusicGen~\cite{Copet2023-sx} and AudioGen~\cite{Kreuk2022-pc} for language model-driven audio generation.

To model the representation similarity between text and audio, one related work is CLAP~\cite{Elizalde2022-ge} which uses contrastive loss. However, we found that using the contrastive loss in CLAP for generation model training did not improve the performance. Instead, we propose a new approach that first computes the representation similarities of audios and texts between different samples. We then minimize the discrepancies between the audios' similarities and the texts' similarities. Additionally, we found that max pooling is better than average pooling for obtaining the sequence level representation from individual time step output.

\section{Representation regularization}
\label{sec:pagestyle}
\begin{figure}[htb]
  \centerline{\includegraphics[width=8.5cm]{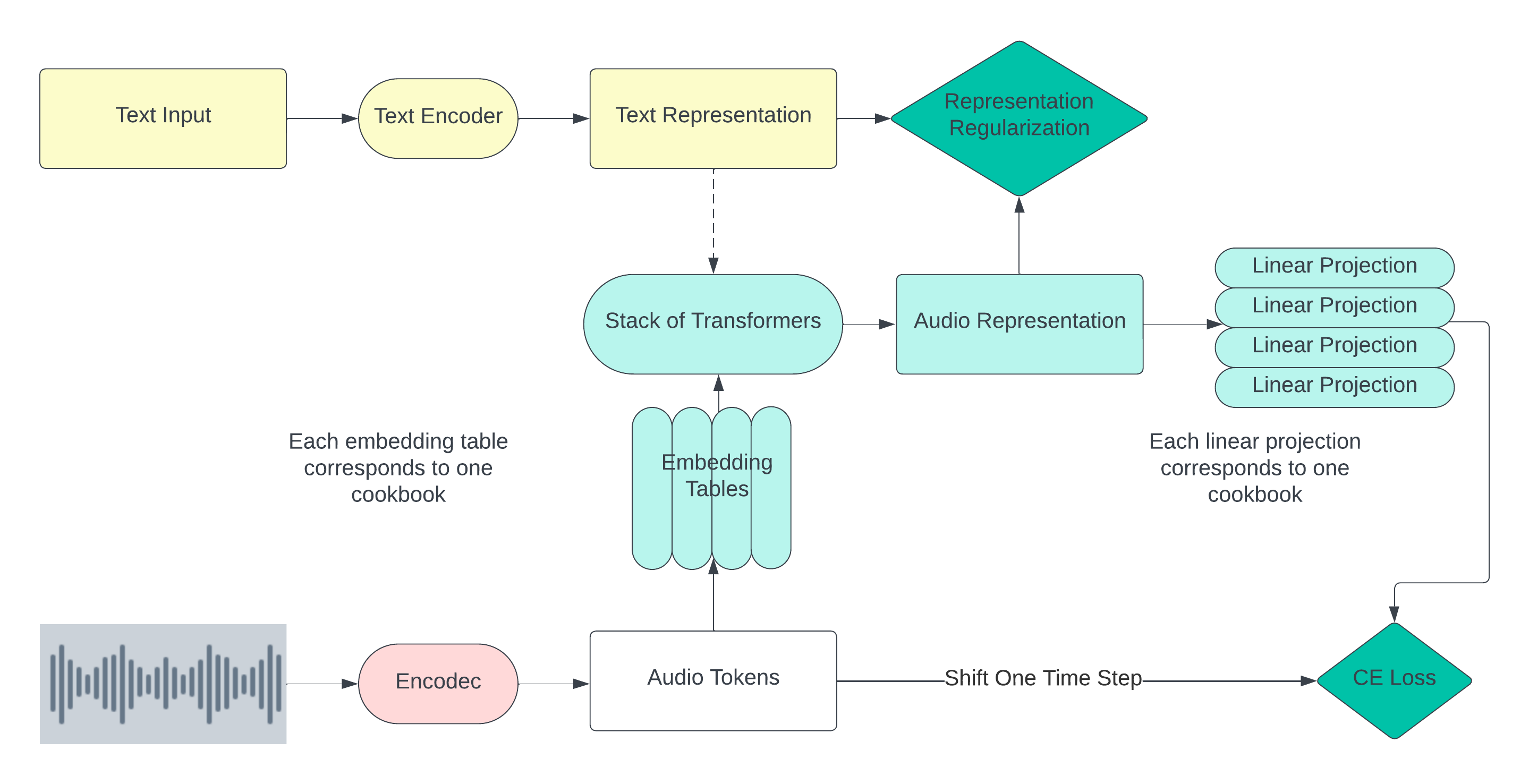}}
\vspace{-3mm}
\caption{Illustration of the language model training with cross entropy loss and representation regularization.}
\label{fig:res}
\vspace{-3mm}
\end{figure}

\subsection{Language model based audio generation}

The language model based audio generation model is composed of several pivotal elements as shown in Fig~\ref{fig:res}. Firstly, it employs a compression model, such as the EnCodec model~\cite{Defossez2022-ub,Zeghidour2022-dr} to encode the raw audio data into a discrete multi-stream sequence of tokens $a_{k,i}$. Here $i\in[1,T_a]$ and $T_a$ is the length of the audio token sequence, while $k\in[1, K]$, indicating the particular codebook indexed as the $k$-th.
Additionally, the model incorporates a pre-trained text encoder, which transforms the text input into a sequence of embedding representations identified as $v_{j}$, where $j\in[1, T_v]$, $T_v$ corresponds to the length of the sequence containing text embedding representations. Lastly, there is a language model component that is a stack of Transformer layers. The language model leverages both the text embedding representation and the preceding audio tokens to generate the probability distribution for the subsequent audio token as $p_\theta(a_{k,i + 1} |a_{k,1},...,a_{k,i},v_{1},...,v_{T_v})$. To render audio generation more manageable, the generation of multi-stream audio tokens is trained in parallel, resulting in a substantial reduction in the effective sequence length during model training. The loss for the language model is the sum of the cross entropy loss for each stream $k$.
\begin{equation}
\small
L_{cond} = -\sum_{k=1}^{K}\sum_{i=1}^{T_a}log(p_\theta(a_{k,i + 1} |a_{k,1},...,a_{k,i},v_{1},...,v_{T_v}))
\label{eq:ce_loss}
\end{equation}

\vspace{-3mm}
\subsection{Representation regularization}
However, the cross entropy loss in language model lacks explicit mechanism to enforce the audio token prediction align with the provided text conditions. Furthermore, the correlation between text and audio gets even loosen as the classifier-free guidance (CFG) method~\cite{Ho2022-wm,Kreuk2022-pc,Copet2023-sx} is used in the training to regulate the balance between sample quality and diversity. Employing CFG involves training the language model both conditionally and unconditionally. Similar to AudioGen~\cite{Kreuk2022-pc}, 10$\%$ of the training samples have their accompanying text omitted during language model training. In unconditional situation, the loss is simply 
\begin{equation}
\small
L_{uncond} = -\sum_{k=1}^{K}\sum_{i=1}^{T_a}log(p_\theta(a_{k,i + 1} |a_{k,1},...,a_{k,i}))
\label{eq:ce_loss}
\end{equation}

In this work, the proposed representation regularization strengthens the correlation between audio representation and text representation while still maintains the effects of CFG method to train the language model unconditionally on text. Given a batch of training samples, a pooling method $F$ is used to get the text sequence representation as $T^b = \mathrm{F}(v_{1}^b,...,v_{T_v}^b)$ and audio sequence representation as $A^b = \mathrm{F}(u_{1}^b,...,u_{T_a}^b)$ for the particular sample $b$ in the batch. In our experiments, the max pooling achieved the best results.

Rather than directly mapping the text and audio representations to the same space and maximizing the similarity between audio and text as CLAP~\cite{Elizalde2022-ge}, we propose to minimize discrepancies in audio and text similarity compared to other samples within the same training batch as follows:
\begin{equation}
\small
T^{b,\hat{b}} = \frac{T^b*T^{\hat{b}}}{||T^b||||T^{\hat{b}}||} 
\label{eq:T}
\end{equation}

\begin{equation}
\small
A^{b,\hat{b}} = \frac{A^b*A^{\hat{b}}}{||A^b||||A^{\hat{b}}||}
\label{eq:A}
\end{equation}

\begin{equation}
\small
L_{rr} = \frac{\sum_{b!=\hat{b}}(T^{b,\hat{b}} - A^{b,\hat{b}})^2}{B * (B -1)} 
\label{eq:rr}
\end{equation}

Here $T^{b,\hat{b}}$ denotes the representation similarity between text inputs in sample $b$ and $\hat{b}$. And $A^{b,\hat{b}}$ denotes the representation similarity between audio in sample $b$ and $\hat{b}$. $B$ is the batch size. The $L_{rr}$ enforces the text and audio in one sample have the same differences regarding to the other samples.

In this study, the proposed representation regularization is exclusively applied during the CFG phase. The complete model training loss is defined as follows:

\begin{equation}
L =
\begin{cases}
L_{uncond} + \lambda L_{rr} & \quad \text{if CFG is utilized}\\
L_{cond} & \quad \text{if CFG is not used}
\end{cases}
\end{equation}
Here, $\lambda$ represents the weighting factor for the representation regularization. Note that representation regularization is only employed during regular training steps when CFG is in use. We also conducted experiments involving representation regularization in non-CFG scenarios; however, these experiments did not yield improvements in objective metrics. We believe the degradation may be attributed to the fact that representation regularization has the potential to hinder language model learning by copying the text representation from cross-attention as the audio representation in non-CFG.

\section{Experiments}
\label{sec:experiments}
In this work, we use two sets of experiments including the sound effects generation and the music generation to verify the effectiveness of proposed methods.
\vspace{-3mm}
\subsection{Datasets}
\label{ssec:datasets}
In music generation, we utilize a total of 20K hours of licensed music which comprises an internal compilation of 10K music tracks of high quality, and 390k instrument-only music tracks from the ShutterStock\footnote{www.shutterstock.com/music} and Pond5\footnote{www.pond5.com}. All datasets are full-length music with 32 kHz sampling rate, accompanied by comprehensive metadata such as textual descriptions, genre categorizations, BPM, and tags. Our evaluation uses the MusicCaps benchmark~\cite{Agostinelli2023-rc}. The MusicCaps benchmark comprises 5.5K samples including a subset of 1K samples balanced across various genres. We report objective metrics on the unbalanced subset as~\cite{Copet2023-sx}.

For sound effect model training, a dataset encompassing 4k hours of training data is employed. This dataset incorporates resources like AudioSet~\cite{Gemmeke2017-ro}, BBC sound effects\footnote{https://sound-effects.bbcrewind.co.uk/}, AudioCaps\cite{Kim2019-zh}, Clotho v2~\cite{Drossos2020-lw}, VGG-Sound~\cite{Chen2020-bi}, FSD50K~\cite{Fonseca2022-ht} and Free To Use Sounds\footnote{https://www.freetousesounds.com/all-in-one-bundle/}. All audio files are sampled at a rate of 16kHz. We adopt a preprocessing methodology akin to~\cite{Kreuk2022-pc} for textual descriptions. To begin, we utilize multi-label annotations from datasets such as AudioSet, VGG-Sound, FSD50K. Pseudo-sentences are constructed by concatenating lists of tags linked with audio samples. Subsequently, we eliminate stop words and numbers, and lemmatize natural language captions available in datasets including AudioCaps, Clotho v2, Free To Use Sounds, and BBC Sound Effects. Lastly, samples containing the term "speech" in their tag or caption are filtered out, given that speech predominates in the data.

\vspace{-3mm}
\subsection{Setup}
\label{ssec:setup}
Our approach involves a non-causal five-layer EnCodec model tailored for music generation, operating at 32 kHz for monophonic music, and 16 kHz for sound effects generation. These EnCodec models maintain a frame rate of 50 Hz, commencing with an initial hidden size of 64, which doubles across the model's five layers. Embeddings are subjected to quantization using an RVQ comprising four quantizers, each featuring a codebook size of 2048. These EnCodec models are trained using the same audio data as those in the language model training.

The transformer models used in this work have 300M parameters. To enhance efficiency with long sequences, we employ memory-efficient Flash attention~\cite{Dao2022-cy} from the xFormers package~\cite{Lefaudeux2021-tn}, improving both speed and memory utilization. For ablations, we consistently employ the sound effects generation model setup. For music generation model training, 30-second audio segments are used, randomly sampled from the complete track. In sound effects generation training, 10-second audio clips are used. Model training spans 100K steps, utilizing the AdamW optimizer~\cite{Loshchilov2017-hk}, a batch size of 192 examples, $\beta_1 = 0.9$, $\beta_2 = 0.95$, a decoupled weight decay of 0.1, and gradient clipping of 1.0. A cosine learning rate schedule is employed, with a warmup of 4k steps. Furthermore, an exponential moving average is applied, characterized by a decay factor of 0.99. The model training employs the mixed precision with Fully Sharded Data Parallel (FSDP) bfloat16. We used 16 GPUs and 32 GPUs for sound effects generation and music generation training, respectively.  In the sampling process for inference, we adopt top-k sampling~\cite{Fan2018-oo}, retaining the top 250 tokens and applying a temperature of 1.0.

\vspace{-3mm}
\subsection{Ablation Study}
\label{ssec:ablation_study}
Table~\ref{tab:ablation} presents the results of the ablation study conducted on the sound effects generation model using the AudioCaps dataset. The optimal model was trained with representation regularization based on max pooling, employing a weight parameter of $\lambda=3.0$ and allocating $10\%$ of the training data for CFG training. In contrast, the use of average pooling-based sequence representation regularization did not demonstrate any improvement over the baseline. Furthermore, Table~\ref{tab:ablation} reaffirms the significant role of CFG training in reducing both FAD and KL scores.

\begin{table}[!h]
\begin{center}
\begin{tabular}[\linewidth]{ c c c | c c c} 
 \toprule
 pool & CFG & $\lambda$ & FAD($\downarrow$) & KL($\downarrow$) & CLAP($\uparrow$) \\  
 \midrule
 max & 0.1 & 3& \bf{1.43}&\bf{1.57}&\bf{0.31}  \\
 \midrule
max & 0.1 & 4 & 1.44&1.58&0.30 \\
max & 0.1 & 2 & 1.56&1.57&0.31 \\
max & 0.1 & 1 & 1.58&1.61&0.30 \\
\midrule
- & 0.2 & 0 & 1.56 & 1.60 & 0.30 \\
- & 0.1 & 0 & 1.52&1.60&0.30 \\
- & 0.0 & 0 & 1.69&1.58&0.30 \\
\midrule
 max & 0.2 & 3& 1.59&1.64&0.30  \\
average & 0.1 & 3 & 1.54&1.59&0.30 \\
\bottomrule
\end{tabular}
\end{center}
\vspace{-5mm}
\caption{Ablation study using sound effects generation based on AudioCaps. The column `pool' denotes the pooling method to get the sequence level representation for both audio and text representation. `CFG' column gives the ratio of using CFG in training. `$\lambda$' represents the weight used in representation regularization.}
\label{tab:ablation} 
\end{table}

\vspace{-5mm}
\subsection{Music Generation}
\label{ssec:Music generation}
Table~\ref{table:music} gives the objective metrics on the MusicCaps data. We report the original metrics for MuiscLM, Noise2Music and MusicGen 1.5B model without melody. Notably, the introduction of the proposed representation regularization results in enhancements across all metrics. Our 300M parameter model, which incorporates representation regularization, surpasses the performance of the MusicGen 1.5B parameter model in terms of FAD and CLAP.

\begin{table}[!h]
\begin{center}
\begin{tabular}[\linewidth]{c c c c} 
 \toprule
 Methods & FAD($\downarrow$) & KL($\downarrow$) & CLAP($\uparrow$) \\  
 \midrule
 MusicLM~\cite{Agostinelli2023-rc} & 4.0 & - & - \\
 Noise2Music\cite{Huang2023-yn} & 2.1 & - & - \\
 MusicGen 1.5B\cite{Copet2023-sx} &5.0  &1.31  &0.28  \\
  \midrule
 ours 300M w/o rr &5.28&1.36&0.30  \\
 ours 300M w/ rr & 4.83&1.32&0.31 \\
\bottomrule
\end{tabular}
\end{center}
\vspace{-5mm}
\caption{Music generation using MusicCaps. 'w/ rr' and 'w/o rr' mean with and without represenation regularization, respectively.}
\label{table:music} 
\end{table}

\vspace{-5mm}
\subsection{Sound Effects Generation}
\label{ssec:sound generation}
The sound effects generation results on AudioCaps are shown in Table~\ref{table:sound}. The trend is the same as the music generation experiments. The representation regularization improves the model performance on FAD, KL and CLAP. The results of AudioGen is referring to the github\footnote{https://github.com/facebookresearch/audiocraft/blob/main/model\_cards}.

\begin{table}[!h]
\begin{center}
\begin{tabular}[\linewidth]{c c c c} 
 \toprule
 Methods & FAD($\downarrow$) & KL($\downarrow$) & CLAP($\uparrow$) \\  
 \midrule
 AudioGen~\cite{Kreuk2022-pc} & 1.77 & 1.58 & 0.30 \\
  \midrule
 ours w/o rr &1.52&1.60&0.30  \\
 ours w/ rr & 1.43&1.57&0.31 \\
 
\bottomrule
\end{tabular}
\end{center}
\vspace{-5mm}
\caption{Sound effects generation using AudioCaps. 'w/ rr' and 'w/o rr' mean with and without represenation regularization, respectively.}
\label{table:sound} 
\end{table}

\vspace{-5mm}
\subsection{Human preference evaluation}
\label{ssec:human}
Table~\ref{table:human} gives the subjective metrics for the sound and music generation models. Our subjective evaluation employed a blind pairwise comparison test, where evaluators were presented with two samples generated by distinct models, all based on the same text prompt. This comparison was conducted across a set of 20 text prompts, and eight human evaluators were tasked with determining their preference for the sample they believed exhibited better quality and better alignment with the provided prompt in each pair.

Notably, both music and sound effects generation, when incorporating representation regularization, garnered higher user preference ratings. A possible explanation for the more significant trend in the sound effects generation is that music tends to be more abstract than sound effects. Consequently, any discrepancies in alignment with the provided text may not be as readily apparent to human evaluators.

\begin{table}[!h]
\begin{center}
\begin{tabular}[\linewidth]{c c c } 
 \toprule
 Methods & music & sound effects \\  
 \midrule
 ours w/o rr &48$\%$&33$\%$  \\
 ours w/ rr &52$\%$&67$\%$ \\
\bottomrule
\end{tabular}
\end{center}
\vspace{-5mm}
\caption{Human preference evaluation}
\label{table:human} 
\end{table}

\vspace{-5mm}
\section{conclusion}
This paper has introduced representation regularization to improve controllability over audio generation by prioritizing alignment between audio and text representations during model training. The proposed method integrated the audio and text similarity regularization, particularly during the classifier-free guidance (CFG) phase, wherein the text condition is excluded from cross attention during language model training. The experimental results, conducted across various audio and music generation tasks, demonstrate that the proposed representation regularization has led to improvements in objective metrics for both audio and music generation. Moreover, these improvements have translated into a noticeable enhancement in human perception regarding audio generation quality and alignment. 



\vfill\pagebreak

\bibliographystyle{IEEEbib}
\small
\bibliography{sample}

\begin{thebibliography}{10}

\bibitem{Rombach2022-yo}
Robin Rombach, Andreas Blattmann, et~al.,
\newblock ``High-resolution image synthesis with latent diffusion models,''
\newblock in {\em CVPR}, 2022.

\bibitem{Ramesh2022-sn}
Aditya Ramesh, Prafulla Dhariwal, et~al.,
\newblock ``Hierarchical {Text-Conditional} image generation with {CLIP}
  latents,''
\newblock {\em arXiv}, 2022.

\bibitem{Song2020-go}
Yang Song, Jascha Sohl-Dickstein, et~al.,
\newblock ``{Score-Based} generative modeling through stochastic differential
  equations,''
\newblock {\em arXiv}, 2020.

\bibitem{Liu2023-uz}
Haohe Liu, Qiao Tian, et~al.,
\newblock ``{AudioLDM} 2: Learning holistic audio generation with
  self-supervised pretraining,''
\newblock {\em arXiv}, Aug. 2023.

\bibitem{Liu2023-bg}
Haohe Liu, Zehua Chen, et~al.,
\newblock ``{AudioLDM}: {Text-to-Audio} generation with latent diffusion
  models,''
\newblock {\em arXiv}, 2023.

\bibitem{Kreuk2022-pc}
Felix Kreuk, Gabriel Synnaeve, et~al.,
\newblock ``{AudioGen}: Textually guided audio generation,''
\newblock {\em arXiv}, 2022.

\bibitem{Agostinelli2023-rc}
Andrea Agostinelli, Timo~I Denk, et~al.,
\newblock ``{MusicLM}: Generating music from text,''
\newblock {\em arXiv}, 2023.

\bibitem{Copet2023-sx}
Jade Copet, Felix Kreuk, et~al.,
\newblock ``Simple and controllable music generation,''
\newblock {\em arXiv}, 2023.

\bibitem{Le2023-fx}
Matthew Le, Apoorv Vyas, et~al.,
\newblock ``Voicebox: {Text-Guided} multilingual universal speech generation at
  scale,''
\newblock {\em arXiv}, 2023.

\bibitem{Lam2023-lu}
Max W~Y Lam, Qiao Tian, et~al.,
\newblock ``Efficient neural music generation,''
\newblock {\em arXiv}, 2023.

\bibitem{Dhariwal2021-tg}
Prafulla Dhariwal and Alexander Nichol,
\newblock ``Diffusion models beat gans on image synthesis,''
\newblock {\em Adv. Neural Inf. Process. Syst.}, 2021.

\bibitem{Ho2020-pp}
Jonathan Ho, Ajay Jain, and Pieter Abbeel,
\newblock ``Denoising diffusion probabilistic models,''
\newblock {\em Adv. Neural Inf. Process. Syst.}, 2020.

\bibitem{Kingma2021-zs}
Diederik Kingma, Tim Salimans, et~al.,
\newblock ``Variational diffusion models,''
\newblock {\em Adv. Neural Inf. Process. Syst.}, 2021.

\bibitem{Huang2022-sc}
Rongjie Huang, Max W~Y Lam, et~al.,
\newblock ``{FastDiff}: A fast conditional diffusion model for {High-Quality}
  speech synthesis,''
\newblock {\em arXiv}, 2022.

\bibitem{Kim2022-xl}
Sungwon Kim, Heeseung Kim, and Sungroh Yoon,
\newblock ``{Guided-TTS} 2: A diffusion model for high-quality adaptive
  {Text-to-Speech} with untranscribed data,''
\newblock {\em arXiv}, 2022.

\bibitem{Shen2023-wp}
Kai Shen, Zeqian Ju, et~al.,
\newblock ``{NaturalSpeech} 2: Latent diffusion models are natural and
  {Zero-Shot} speech and singing synthesizers,''
\newblock {\em arXiv}, 2023.

\bibitem{Huang2023-zw}
Rongjie Huang, Jiawei Huang, et~al.,
\newblock ``{Make-An-Audio}: {Text-To-Audio} generation with {Prompt-Enhanced}
  diffusion models,''
\newblock {\em arXiv}, 2023.

\bibitem{Schneider2023-ju}
Flavio Schneider, Zhijing Jin, and Bernhard Sch{\"o}lkopf,
\newblock ``Mo{\^u}sai: {Text-to-Music} generation with {Long-Context} latent
  diffusion,''
\newblock {\em arXiv}, 2023.

\bibitem{Vaswani2017-mw}
Ashish Vaswani, Noam Shazeer, et~al.,
\newblock ``Attention is all you need,''
\newblock {\em Adv. Neural Inf. Process. Syst.}, 2017.

\bibitem{Borsos2023-xo}
Zal{\'a}n Borsos, Rapha{\"e}l Marinier, et~al.,
\newblock ``{AudioLM}: A language modeling approach to audio generation,''
\newblock {\em IEEE/ACM Transactions on Audio, Speech, and Language
  Processing}, 2023.

\bibitem{Dunbar2021-os}
Ewan Dunbar, Mathieu Bernard, et~al.,
\newblock ``The zero resource speech challenge 2021: Spoken language
  modelling,''
\newblock {\em arXiv}, 2021.

\bibitem{Lakhotia2021-ir}
Kushal Lakhotia, Eugene Kharitonov, et~al.,
\newblock ``On generative spoken language modeling from raw audio,''
\newblock {\em Transactions of the Association for Computational Linguistics},
  2021.

\bibitem{Zeghidour2022-dr}
Neil Zeghidour, Alejandro Luebs, et~al.,
\newblock ``{SoundStream}: An {End-to-End} neural audio codec,''
\newblock {\em IEEE/ACM Transactions on Audio, Speech, and Language
  Processing}, 2022.

\bibitem{Defossez2022-ck}
Alexandre D{\'e}fossez, Jade Copet, et~al.,
\newblock ``High fidelity neural audio compression,''
\newblock {\em arXiv}, 2022.

\bibitem{Raffel2019-ba}
Colin Raffel, Noam Shazeer, et~al.,
\newblock ``Exploring the limits of transfer learning with a unified
  {Text-to-Text} transformer,''
\newblock {\em arXiv}, 2019.

\bibitem{Ho2022-wm}
Jonathan Ho and Tim Salimans,
\newblock ``{Classifier-Free} diffusion guidance,''
\newblock {\em arXiv}, 2022.

\bibitem{Hershey2017-fk}
Shawn Hershey, Sourish Chaudhuri, et~al.,
\newblock ``{CNN} architectures for large-scale audio classification,''
\newblock in {\em ICASSP}, 2017.

\bibitem{Koutini2021-fx}
Khaled Koutini, Jan Schl{\"u}ter, et~al.,
\newblock ``Efficient training of audio transformers with patchout,''
\newblock {\em arXiv}, 2021.

\bibitem{Elizalde2022-ge}
Benjamin Elizalde, Soham Deshmukh, Mahmoud Al~Ismail, and Huaming Wang,
\newblock ``{CLAP}: Learning audio concepts from natural language
  supervision,''
\newblock {\em arXiv}, 2022.

\bibitem{Defossez2022-ub}
Alexandre D{\'e}fossez, Jade Copet, Gabriel Synnaeve, and Yossi Adi,
\newblock ``High fidelity neural audio compression,''
\newblock {\em arXiv}, 2022.

\bibitem{Gemmeke2017-ro}
Jort~F Gemmeke, Daniel P~W Ellis, et~al.,
\newblock ``Audio set: An ontology and human-labeled dataset for audio
  events,''
\newblock in {\em ICASSP}, 2017.

\bibitem{Kim2019-zh}
Chris~Dongjoo Kim, Byeongchang Kim, et~al.,
\newblock ``{{A}udio{C}aps}: Generating captions for audios in the wild,''
\newblock in {\em NAACL}, 2019.

\bibitem{Drossos2020-lw}
Konstantinos Drossos, Samuel Lipping, and Tuomas Virtanen,
\newblock ``Clotho: an audio captioning dataset,''
\newblock in {\em ICASSP}, 2020.

\bibitem{Chen2020-bi}
Honglie Chen, Weidi Xie, et~al.,
\newblock ``Vggsound: A {Large-Scale} {Audio-Visual} dataset,''
\newblock in {\em ICASSP}, 2020.

\bibitem{Fonseca2022-ht}
Eduardo Fonseca, Xavier Favory, et~al.,
\newblock ``{FSD50K}: An open dataset of {Human-Labeled} sound events,''
\newblock {\em IEEE/ACM Transactions on Audio, Speech, and Language
  Processing}, 2022.

\bibitem{Dao2022-cy}
Tri Dao, Daniel~Y Fu, et~al.,
\newblock ``{FlashAttention}: Fast and memory-efficient exact attention with
  {IO-awareness},''
\newblock {\em arXiv}, 2022.

\bibitem{Lefaudeux2021-tn}
Benjamin Lefaudeux, Francisco Massa, et~al.,
\newblock ``xformers: A modular and hackable transformer modelling library,''
  2021.

\bibitem{Loshchilov2017-hk}
Ilya Loshchilov and Frank Hutter,
\newblock ``Decoupled weight decay regularization,''
\newblock {\em arXiv}, 2017.

\bibitem{Fan2018-oo}
Angela Fan, Mike Lewis, and Yann Dauphin,
\newblock ``Hierarchical neural story generation,''
\newblock {\em arXiv}, 2018.

\bibitem{Huang2023-yn}
Qingqing Huang, Daniel~S Park, et~al.,
\newblock ``{Noise2Music}: Text-conditioned music generation with diffusion
  models,''
\newblock {\em arXiv}, 2023.

\end{thebibliography}

\end{document}